\documentclass[fp,twocolumn]{jpsj3}

\usepackage[dvipdfmx]{graphicx}
\usepackage{pdfpages}
\usepackage{dcolumn}
\usepackage{bm}
\usepackage{txfonts}
\usepackage{color}
\usepackage{amssymb}

\title{Strain-Induced Landau Levels of Majorana Fermions in an Anisotropically Interacting Kitaev Model on a Honeycomb Lattice}

\author{Takuto Yamada and Sei-ichiro Suga}
\inst{Graduate School of Engineering, University of Hyogo, Himeji 671-2280, Japan} 

\abst{The energy structure of an anisotropically interacting Kitaev model on a honeycomb lattice under triaxial strain is investigated. A numerical calculation shows that quantized states appear in the low-energy region, even when the anisotropy of the interaction is rather strong. Their energies are proportional to the square root of the quantum number and the quantized state at zero energy appears only on one sublattice. These findings indicate the emergence of the strain-induced Landau levels of Majorana fermions, which is also confirmed by an analytical calculation. These Landau levels are stable, when the direction of triaxial strain is slightly changed from the bond direction.}

\begin{document}
\maketitle

\section{\label{sec:level1}Introduction}
The Kitaev model is an $S=1/2$ quantum spin model that has bond-dependent Ising-type interactions on a honeycomb lattice \cite{Kitaev}, called Kitaev interactions. 
A Majorana representation of the spin operators was shown that this model is described by noninteracting itinerant Majorana fermions coupled with ${\mathbb Z}_2$ gauge fluxes and that the ground state is in the flux-free sector \cite{Kitaev}. 
Furthermore, the ground state was shown to be a quantum spin liquid \cite{Baskaran}. 
In the ground state, the low-lying dispersion is described by the type of Dirac fermions.   
Fascinating properties related with Majorana fermions have been revealed by intensive theoretical studies. 
Materials exhibiting Kitaev interactions, called Kitaev candidate materials have been found, including ${\rm A_2IrO_3 \; (A=Na, Ir)}$ \cite{Chaloupka,Singh2,Comin,Foyevtsova,Sizyuku,Katukuri,Yamaji2014,Chun,Winter2016}, $\alpha$-${\rm RuCl_3}$ \cite{Plump,Kubota,Majumder,Sandilands,HSKim2016,Yadav,Sinn,WangPRB,Winter2016}, and ${\rm H_3LiIr_2O_6}$ \cite{Kitagawa}. 
The behavior caused by Majorana fermions in these materials has been studied using various methods \cite{Hermanns,Knolle2019,Takagi,Motome2020,Trebst}. 
In their results, half-integer thermal quantum Hall effect can be a conclusive evidence for the emergent itinerant Majorana fermions. This phenomenon has been first pointed out theoretically \cite{Kitaev} and then observed experimentally in $\alpha$-${\rm RuCl_3}$ \cite{Kasahara1,Kasahara2}.

Since Majorana fermions are charge-neutral particles acting as their own antiparticles, they are difficult to interact directly to electromagnetic fields.  
Strain fields can induce an pseudovector potential for Dirac fermions, which has opposite signs at two Dirac points due to time-reversal symmetry \cite{Suzuura,Vozmediano}. Experiments on strained graphene  \cite{Levy,Lu,Liu,Nigge} and artificial  strained graphene \cite{Gomes,Rechtsman} have revealed a strong pseudomagnetic field in the range of $10$ T--$100$ T and the presence of Landau levels. 
The strain-induced pseudomagnetic field is considered to interact directly with itinerant Majorana fermions. 
Indeed, numerical calculations have shown that the Landau levels of itinerant Majorana fermions emerge in the low-energy region of the isotropically interacting Kitaev model under triaxial strain \cite{Rachel}. The related phenomena with these Landau levels have been also investigated theoretically \cite{Rachel,Perreault,Agarwala,Fremling}. 
Thus, the phenomena related to the strain-induced Landau levels in the Kitaev candidate materials can be a hallmark of itinerant Majorana fermions.

The {\it ab-initio} calculations for the Kitaev candidate materials have argued that the Kitaev interaction is dominant and anisotropic \cite{Yamaji2014,Sizyuku,HSKim2016}. 
Furthermore, it was reported in Ref. \cite{Rachel} that a two-flux excitation induces bound states between a sequence of the Landau levels of Majorana fermions. Thus, to observe the clearly identifiable Landau levels, it is indispensable to investigate them in the flux-free sector.  
In this study, we explore the energy structure of the anisotropically interacting Kitaev model on a honeycomb lattice under triaxial strain. We focus on the parameter space spanned by the three Kitaev interactions where the itinerant Majorana fermions exhibit a gapless dispersion relation in the absence of a strain field. We first confirm that the ground state is in the flux-free sector for given strain strength and the system size, when the Kitaev interactions are changed systematically. We then demonstrate that the clearly identifiable multiple Landau levels of Majorana fermions emerge in the low-energy region through a numerical calculation. The results are confirmed also by an analytical calculation. We further find that these Landau levels are stable, when the direction of triaxial strain is slightly changed.

The rest of the paper is organized as follows.
Section \ref{sec:level2} outlines the transformation of the Kitaev model for the numerical calculation using a singular-value decomposition method. We then determine the ${\mathbb Z}_2$ gauge-flux sector of the ground state for given strain strength, the system size, and the Kitaev interactions numerically.  
Section \ref{sec:level3} presents the numerical results for the local density of states (LDOS) of the itinerant Majorana fermions; we show the presence of the strain-induced Landau levels of Majorana fermions in the anisotropically interacting Kitaev model. 
These Landau levels are robust against a slight deviation of the strain direction from the bond direction.  
Section \ref{sec:level4} discusses the low-energy states of the system based on the analytical calculation, illustrating results consistent with the numerical outcomes.  
Finally, the study is summarized in Sec. \ref{sec:level5}.

\section{\label{sec:level2}Model and method}
\subsection{\label{sec:level2-1}Formulation for numerical calculations}
The Hamiltonian is described by
\begin{eqnarray}
{\mathcal H}=-\sum_{{\langle jk \rangle}_{x}} J_{jk}^x \sigma_j^x \sigma_k^x - \sum_{{\langle jk \rangle}_{y}} J_{jk}^y \sigma_j^y \sigma_k^y - \sum_{{\langle jk \rangle}_{z}} J_{jk}^z \sigma_j^z \sigma_k^z, 
\label{Ham1}
\end{eqnarray}
where $\sigma_{j}^{\alpha} \; (\alpha= x,y,z)$ is an $\alpha$ component of the Pauli matrix at the $j$ site and $J_{jk}^{\alpha}$ is the coupling constant between the nearest-neighbor atoms on the ${\alpha}$ bond in the honeycomb lattice. 
%
\begin{figure}[thb]
\begin{center}
\includegraphics[width=0.9\hsize]{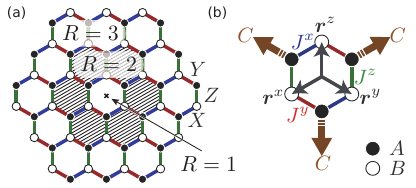}
\vspace{-5mm}
\end{center}
\caption{(Color online) \label{fig1}(a) Unstrained honeycomb flakes expressed by $R$: $R=1$ is a central hexagon (a cross denotes its center.), $R=2$ consists of a central hexagon and six surrounding hexagons, $R=3$ consists of the $R=2$ and twelve surrounding hexagons, and so on. $R$ honeycomb flake includes $2N=6R^2$ spins. 
The $A$ and $B$ sublattices are shown in black and white, respectively. $X$, $Y$, and $Z$ bonds are represented in blue, red, and green, respectively. 
(b) Central hexagon of the unstrained honeycomb lattice. The vectors connect correspondingly the nearest-neighbor sites along the bonds, respectively. Triaxial strain $C$ is represented schematically using three brown arrows.  }
\end{figure}
We use a zigzag-terminated honeycomb lattice with an open boundary condition. The size of the honeycomb flakes is expressed by $R$ [Fig. \ref{fig1}(a)] \cite{Rachel} and the system described by $R$ includes $2N=6R^2$ spins with $N$ being the number of the unit cell. The triaxial strain originates at the center of the central hexagon marked by an cross in Fig. \ref{fig1}(a).
In the unstrained honeycomb lattice, the coupling constants are independent of the site: $J_{jk}^{\alpha}=J^{\alpha}(>0)$. 
When weak triaxial strain is applied as schematically shown in Fig. \ref{fig1}(b), the coupling constant $J_{jk}^{\alpha}$ becomes \cite{Guinea,Amal,Settnes, Rachel} $J_{jk}^{\alpha} \approx J^{\alpha}\left[1-\beta\left(1-|{\boldsymbol r}_j-{\boldsymbol r}_k|/a_0 \right) \right]$, where $\beta$ is the magnetoelastic coupling and $a_0$ is the unstrained bond length. 
The position vector of an atom is given by ${\boldsymbol r_j}={\boldsymbol r^0_{j}}+{\boldsymbol u_j}$, where ${\boldsymbol r^0_{j}=(x^0_{j}, y^0_{j})}$ is the position vector in the unstrained lattice and ${\boldsymbol u}_j$ is the displacement vector expressed as ${\boldsymbol u}_j=\left(C/a_0\right)(2x^0_jy^0_j, {x^0_j}^2-{y^0_j}^2)$, where $C$ is the triaxial strain strength. $J_{jk}^{\alpha}$ must be positive on the whole nearest-neighbor bonds. According to our numerical calculation, this condition is satisfied for $CR \lessapprox 0.3$. We thus set $CR=0.2$ in the following numerical calculation.  
In the honeycomb flakes possessing the same constant $CR$, a scaling holds concerning the honeycomb flake shapes for different $R$ values \cite{Fremling}.

To diagonalize the Hamiltonian, four Majorana fermions, $c_j$ and $b_j^{\alpha}$, are set at each site \cite{Kitaev}, satisfying $\{c_j, c_k \}=2\delta_{jk}, \{c_j, b^{\alpha}_k \}=0$, and $\{b^{\alpha}_j, b^{\beta}_k \}=2\delta_{\alpha\beta} \delta_{jk}$. 
To project the enlarged Hilbert space into the physical Hilbert space, the constraint $c_jb_j^xb_j^yb_j^z=1$ is imposed.  
In this procedure, the spin operator is represented as $\sigma_j^{\alpha}=ic_jb_j^{\alpha}$ and the Hamiltonian reads as ${\mathcal H}_u=i\sum_{\alpha\in\{x,y,z\} }\sum_{{\langle jk \rangle}_{\alpha}} J_{jk}^{\alpha} u_{jk}^{\alpha}c_jc_k$, where $u_{jk}^{\alpha}=ib_j^{\alpha}b_k^{\alpha}$ is a bond operator with an eigenvalue of $\pm1$ and satisfies $[{\mathcal H}_u,\ u_{jk}^{\alpha}]=0$. 
Thus, $u_{jk}^{\alpha}$ is identified with a static ${\mathbb Z}_2$ gauge field between the nearest-neighbor $j$ and $k$ sites on the $\alpha$ bond. 
We then introduce a relevant gauge-flux operator defined as a product of the six ${\mathbb Z}_2$ gauge fields surrounding a hexagon \cite{Kitaev}. The gauge-flux operator commutes with ${\mathcal H}_u$ and its eigenvalue becomes $\pm1$. 
Therefore, the system can be mapped to itinerant Majorana fermions coupled with the ${\mathbb Z}_2$ gauge fluxes on the hexagonal plaquettes.  
For every configurations of the ${\mathbb Z}_2$ gauge fluxes, the Hamiltonian ${\mathcal H}_u$ can be expressed as \cite{Rachel}
\begin{eqnarray}
{\mathcal H}_u=\frac{i}{2}
\begin{pmatrix}
  {\bar{c}_A}^{\mathrm T} & {\bar{c}_B}^{\mathrm T}
\end{pmatrix}
\begin{pmatrix}
      0   & M \\
  -M^{\mathrm T} & 0 \\
\end{pmatrix}
\begin{pmatrix}
  {\bar{c}_A} \\
  {\bar{c}_B} \\
\end{pmatrix},
\label{Ham2}
\end{eqnarray}
where $M_{jk}=J_{jk}^{\alpha} u_{jk}^{\alpha}$ and $\bar{c}_{A(B)}$ is an $N$-component vector representing the itinerant Majorana fermions on the $A(B)$ sublattice. We call the ${\mathbb Z}_2$ gauge-flux having $-1$ `flux'. 
When at least two of the three coupling constants are equal in the unstrained system, the Lieb's theorem \cite{Lieb} states that the exact ground state is in the sector where all the ${\mathbb Z}_2$ gauge fluxes take unity (the flux-free sector) \cite{Kitaev}. The sector where the $n$ gauge fluxes become $-1$ is called the $n$-flux sector.

By using a singular-value decomposition method, we calculate the eigenvalues $\epsilon_{m,\boldsymbol{n}}$ $(m=1,2, \cdots,N)$ and the eigenvectors for a given $n$-flux configuration $\boldsymbol{n}$; then we obtain the LDOS, $\rho_{j,A(B)}(E)$, of the itinerant Majorana fermions on the $A(B)$ sublattice in the $j$-th unit cell.  
The magnetoelastic coupling is set as $\beta=1$ for simplicity. The coupling constants in the unstrained lattice satisfy $J^x+J^y+J^z=J$. We set $J=1$ as the unit of energy. They form the triangle in the parameter space expressed by $J^{x}$, $J^{y}$, and $J^{z}$ [left panel of Fig. \ref{PD}] \cite{Kitaev}, while the central downward triangle enlarged in the right panel of Fig. \ref{PD} represents the gapless phase.

\subsection{\label{sec:level2-2}One-flux gap and ground-state sector}
\begin{figure}[thb]
\begin{center}
\includegraphics[width=0.9\hsize]{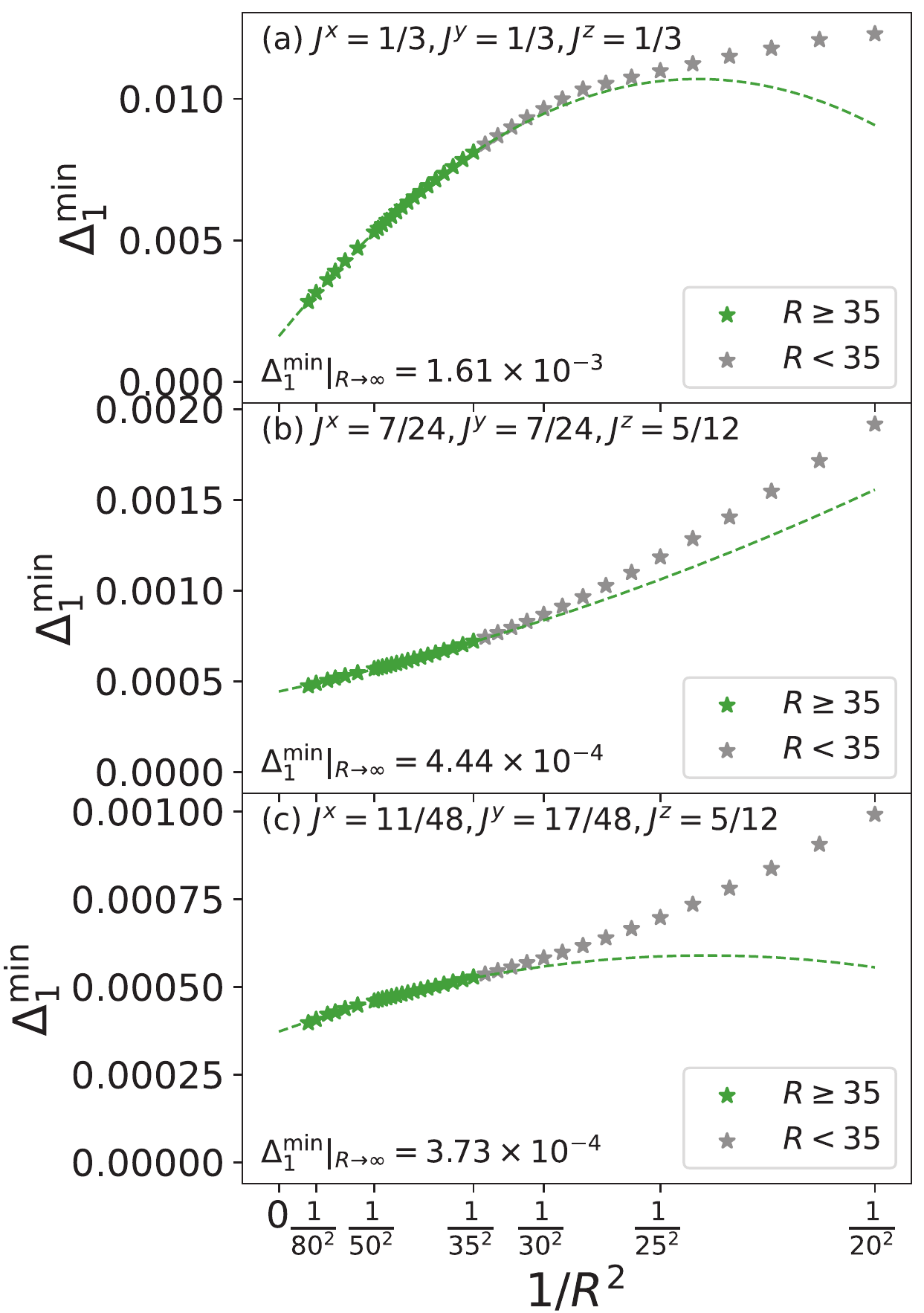}
\vspace{-5mm}
\end{center}
\caption{(Color online) \label{figgap}(a) Minimum one-flux gap ($\Delta_{\boldsymbol 1}^{\rm min}$) for a given $R$. The dashed lines 
represent the polynomial, $\Delta_{\boldsymbol 1}^{\rm min}=aR^{-4}+bR^{-2}+c$, that well describes $\Delta_{\boldsymbol 1}^{\rm min}$ for $R\geq35$ with the constants $a$, $b$, and $c$ having the following values: (a) $-2.95\times10^{3}, 1.04\times10^{1}, 1.61\times10^{-3}$; (b) $6.56\times10^{1}, 2.81\times10^{-1}, 4.44\times10^{-4}$; (c) $-6.77\times10^{1}, 2.42\times10^{-1}, 3.73\times10^{-4}$. 
}
\end{figure}
In the strained honeycomb lattice, the translational invariance is broken, and hence the Lieb's theorem cannot be adopted. Thus, we must confirm whether the ground state is in the flux-free sector for $CR=0.2$. The ground-state energy for a $n$-flux configuration, $\boldsymbol{n}$, is given by $E_{{\rm GS},\boldsymbol{n}}=-\sum_{m}\epsilon_{m,\boldsymbol{n}}$. In the open boundary system, the one-flux state is possible and can be a candidate competing with the flux-free state \cite{Fremling}. 
We calculate the one-flux gap $\Delta_{\boldsymbol 1}=E_{{\rm GS},{\boldsymbol 1}}-E_{{\rm GS},0}$ for all the one-flux configurations at various $R$ up to $90$ for the given $J^x$, $J^y$, and $J^z$. 
Figure \ref{figgap} depicts the typical behavior of the minimum one-flux gap $\Delta_{\boldsymbol 1}^{\rm min}$ for a given $R$. 
When $R\geq35$, $\Delta_{\boldsymbol 1}^{\rm min}$ is well described by the following polynomial: $\Delta_{\boldsymbol 1}^{\rm min}=aR^{-4}+bR^{-2}+c$, where $a, b$, and $c$ are the constants. 
In other words, the system size $R\geq35$ is required to correctly evaluate the extrapolated  $\Delta_{\boldsymbol 1}^{\rm min}$ for $R \rightarrow \infty$. 
The extrapolated $c$ values for $R \rightarrow \infty$ are $1.61\times 10^{-3}$, $4.44\times 10^{-4}$, and $3.73\times 10^{-4}$ in Figs. \ref{figgap}(a)-\ref{figgap}(c), respectively. 
We perform the same calculations for the given coupling constants marked by the black dots in the right panel of Fig. \ref{PD}, finding that all the extrapolated $c$ values for $R \rightarrow \infty$ positive. Thus, we can deduce that the ground state of the anisotropically interacting Kitaev model for $CR=0.2$ is in the flux-free sector. 
In the following numerical calculations, we set $R=60 \; (2N=21600)$ and $C=1/300$. 
We show the LDOS, $\rho_{j, A}(E)$ and  $\rho_{j, B}(E)$, at the site in the central hexagon of the system.

\section{\label{sec:level3}Numerical results for the lattice model}
\subsection{\label{sec:level3-1}Landau levels of itinerant Majorana fermions}
\begin{figure}[thb]
\begin{center}
\includegraphics[width=0.9\hsize]{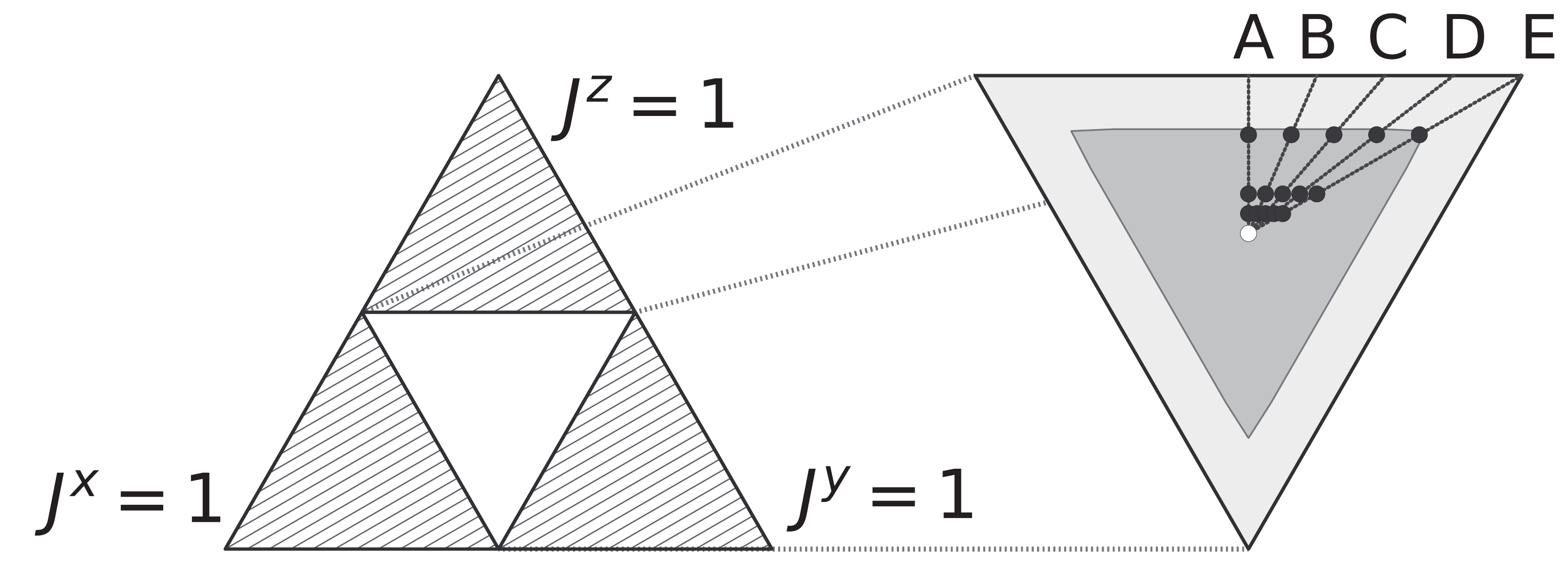}
\end{center}
\caption{(Color online)\label{PD}  Phase diagrams of the unstrained Kitaev model on the plane $J^x+J^y+J^z=1$, where $J^{\alpha} \; (\alpha=x,y,z)$ are the coupling constants of the unstrained system with $J^{\alpha} \geq0$. In the left panel, the inner triangle represents the gapless phase and the three outer triangles are gapped phases. A gapless phase is enlarged in the right panel, where over three Landau levels of Majorana fermions appear in the inner dark shaded region and the sublattice polarization is satisfied for $R=60$ and $C=1/300$. In the outer thin shaded area, one or two peaks appear at and next to $E=0$ in $\rho_{j, A}(E)$, and the sublattice polarization is satisfied. 
}
\end{figure}
\begin{figure*}[thb]
\begin{center}
\includegraphics[width=0.78\hsize]{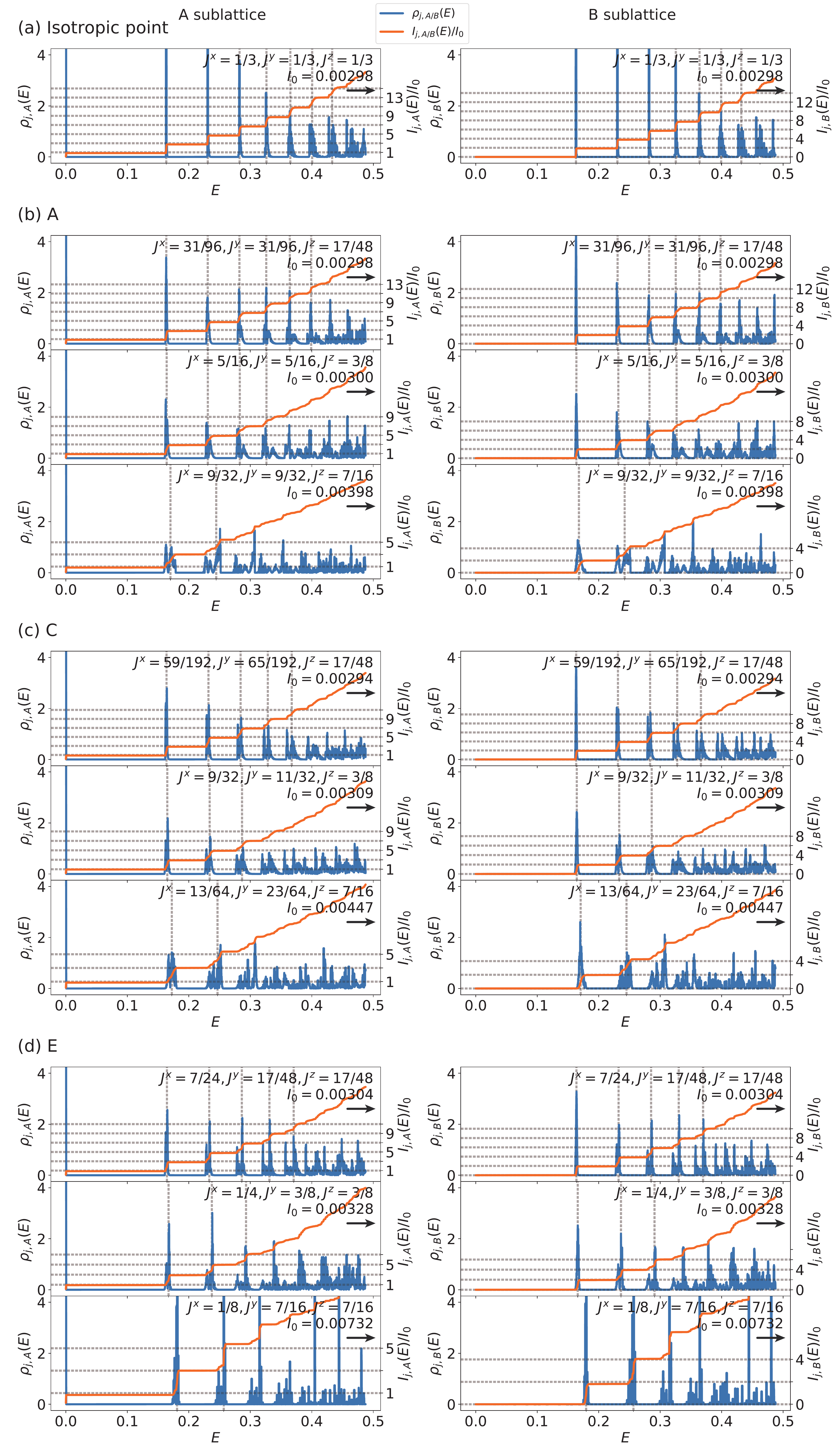}
\hspace{0pc}
\vspace{0pc}
\caption{(Color online)\label{fig2} (a)-(d) Local density of states, $\rho_{j, A/B}(E)$, at the central hexagon of the system and its integral values, $I_{j, A/B}(E)=\int_0^E \rho_{j, A/B}(E') dE'$, for $R=60$ and $C=1/300$. $I_{j, A/B}(E)$ is normalized by numerically evaluated $I_0$ to fit $(2n+1) \; (n=0, 1, 2, \cdots)$ for the $A$ sublattice and $2n \; (n=0, 1, 2, \cdots)$ for the $B$ sublattice. The transverse dotted lines represent $(2n+1)$ for the $A$ sublattice and $2n$ for the $B$ sublattice, respectively.  
 (a) $\rho_{j, A/B}(E)$ and $I_{j, A/B}(E)$ for the isotropic interactions plotted using the small open circle in the right panel of Fig. \ref{PD}; (b)-(d) the coupling constants of the top, middle, and bottom panels correspond to the black dots from close to the center toward the edge along the lines A, C, and E in Fig. \ref{PD}, respectively. 
}
\end{center}
\end{figure*}
\begin{figure*}[htb]
\begin{center}
\includegraphics[width=0.8\hsize]{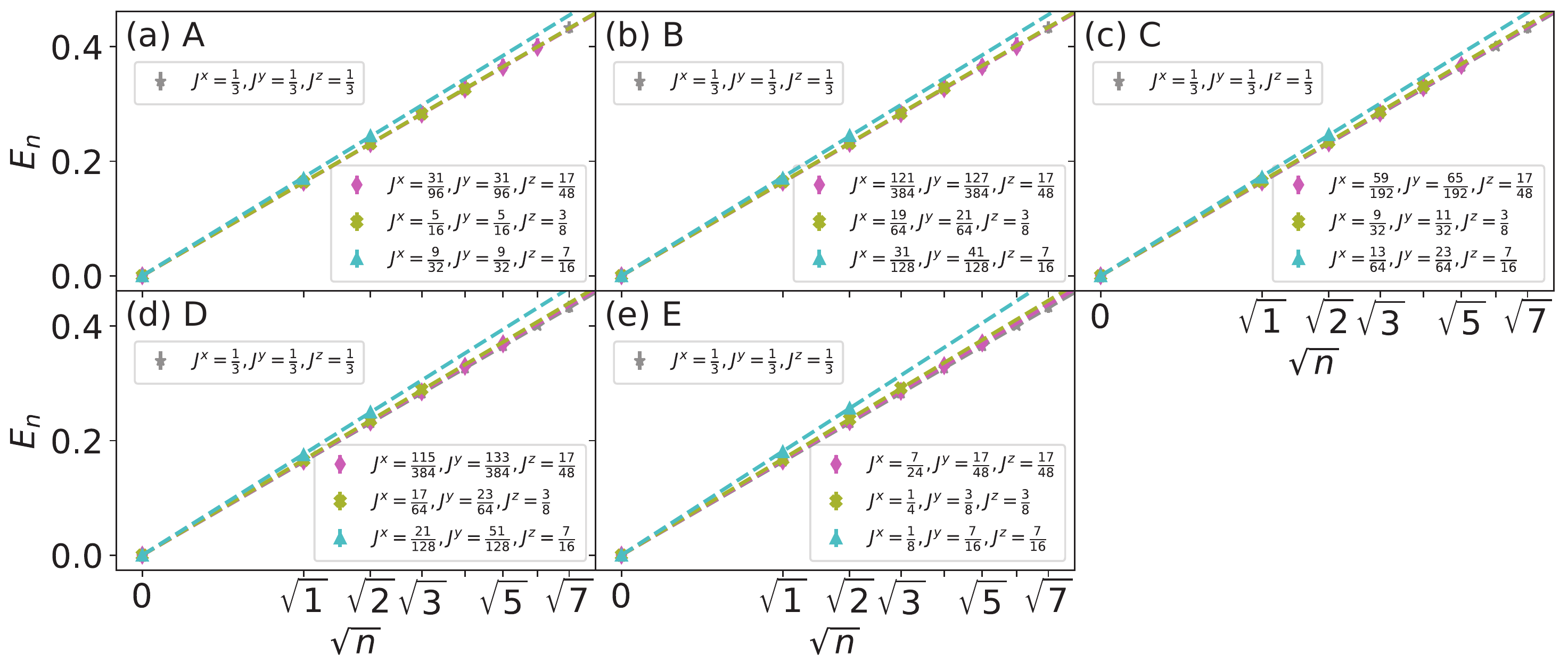}
\hspace{0pc}
\vspace{0pc}
\caption{(Color online)\label{fig3} Peak energies, $E_n \; (n=0, 1, 2, \cdots)$, of  $\rho_{j, A}(E)$ as functions of  $\sqrt{n}$.
The coupling constants in (a)-(e) correspond to the three black dots along the lines A-E in Fig. \ref{PD}, respectively. }
\end{center}
\end{figure*}
Figure \ref{fig2}(a) displays $\rho_{j, A}(E)$ and $\rho_{j, B}(E)$ for the isotropic interactions that are plotted using the small open circle in the right panel of Fig. \ref{PD}. 
Figures \ref{fig2}(b)-\ref{fig2}(d) illustrate the typical results for $\rho_{j, A}(E)$ and $\rho_{j, B}(E)$ for the anisotropic interactions. We also evaluate its integral value: $I_{j, A/B}(E)=\int_0^E \rho_{j, A/B}(E') dE'$. 
The coupling constants of the top, middle, and bottom panels in Figs. \ref{fig2}(b)-\ref{fig2}(d) correspond to the black dots from close to the center toward the edge along the lines A, C, and E [right panel of Fig. \ref{PD}], respectively. 
The left and right panels show the results for the $A$ and $B$ sublattices, respectively.

We find that $I_{j, A/B}(E)$ forms plateaus. The following relations are adopted to explain the pronounced plateaus: $I_{j, A}(E)/I_0=(2n+1) \; (n=0, 1, 2, \cdots)$ for the $A$ sublattice and $I_{j, B}(E)/I_0=2n \; (n=0, 1, 2, \cdots)$ for the $B$ sublattice, the right hand sides of which are represented by the transverse dotted lines in Fig. \ref{fig2}.
The normalization constant $I_0$ is evaluated so as to fit the numerical results. We find that these relations well describe the numerical results.

In the vicinity of the boundary between the pronounced neighboring plateaus,  $\rho_{j, A/B}(E)$ reaches a peak, as indicated by the vertical dashed lines in Figs. \ref{fig2}(a)-\ref{fig2}(d). 
This plateau structure of $I_{j, A/B}(E)$ means that each peak in $\rho_{j, A/B}(E)$ includes the same number of states, which is one of criteria for determining the peak structure in $\rho_{j, A/B}(E)$ to be the Landau level. 
The peak at $E=0$ generally appears only on the $A$ sublattice. This is called the sublattice polarization \cite{Poli}. 
We then plot the peak energies, $E_n \; (n=0, 1, 2, \cdots)$, on the $A$ sublattice (Fig. \ref{fig3}), whose coupling constants correspond to the black dots along the lines A, C, and E [right panel of Fig. \ref{PD}]. As shown in Figs. \ref{fig3}(a), \ref{fig3}(c), and \ref{fig3}(e), $E_n$ satisfies the relation $E_n \propto \sqrt{n}$. Also for the coupling constants at the black dots along the lines B and D, these three features are obtained. 
These findings are characteristic of the Landau levels of gapless Dirac fermions with time-reversal symmetry \cite{Gomes,Amal,Settnes,Uchoa,Fu}. Thus, the itinerant Majorana fermions under triaxial strain are quantized to the Landau levels.

Figures \ref{fig2}(a)-\ref{fig2}(d) indicate that as the system leaves the isotropically interacting point in the right panel of Fig. \ref{PD}, the Landau levels of Majorana fermions are smeared at the higher energies and their number is reduces at the lower energies. 
It is considered that this behavior results from the transfer of the van-Hove-singularity energy to the lower energies as the system leaves the isotropically interacting point, which reduces the energy region of the linear dispersion. 
Within the dark shaded areas on the lines A-E in the right panel of Fig. \ref{PD}, at least three Landau levels of Majorana fermions from $n=0$ appear on the $A$ sublattice, confirming the relation $E_n \propto \sqrt{n}$.     
From the permutation of $J^x$, $J^y$, and $J^z$, there are six equivalent regions in the phase diagram shown in Fig. \ref{PD}. We apply our results to the five additional regions and summarize the results in the right panel of Fig. \ref{PD}. In the dark shaded area, the Landau levels of Majorana fermions emerge. In the outer thin shaded area, instead, one or two peaks at and next to $E=0$ appear in $\rho_{j, A}(E)$, and the sublattice polarization is satisfied. 
We perform the same calculations for $R=40$, $45$, and $50$ while keeping $CR=0.2$. As $R$ increases, the region where the Landau levels of Majorana fermions appear expands toward the boundary between gapless and gapped phases. 
{Therefore, the Landau levels of Majorana fermions are expected to emerge in the whole unstrained gapless phase, when the system becomes large enough.}

\subsection{\label{sec:level3-2}Effects of the direction of triaxial strain}
\begin{figure*}[thb]
\begin{center}
\includegraphics[width=0.76\hsize]{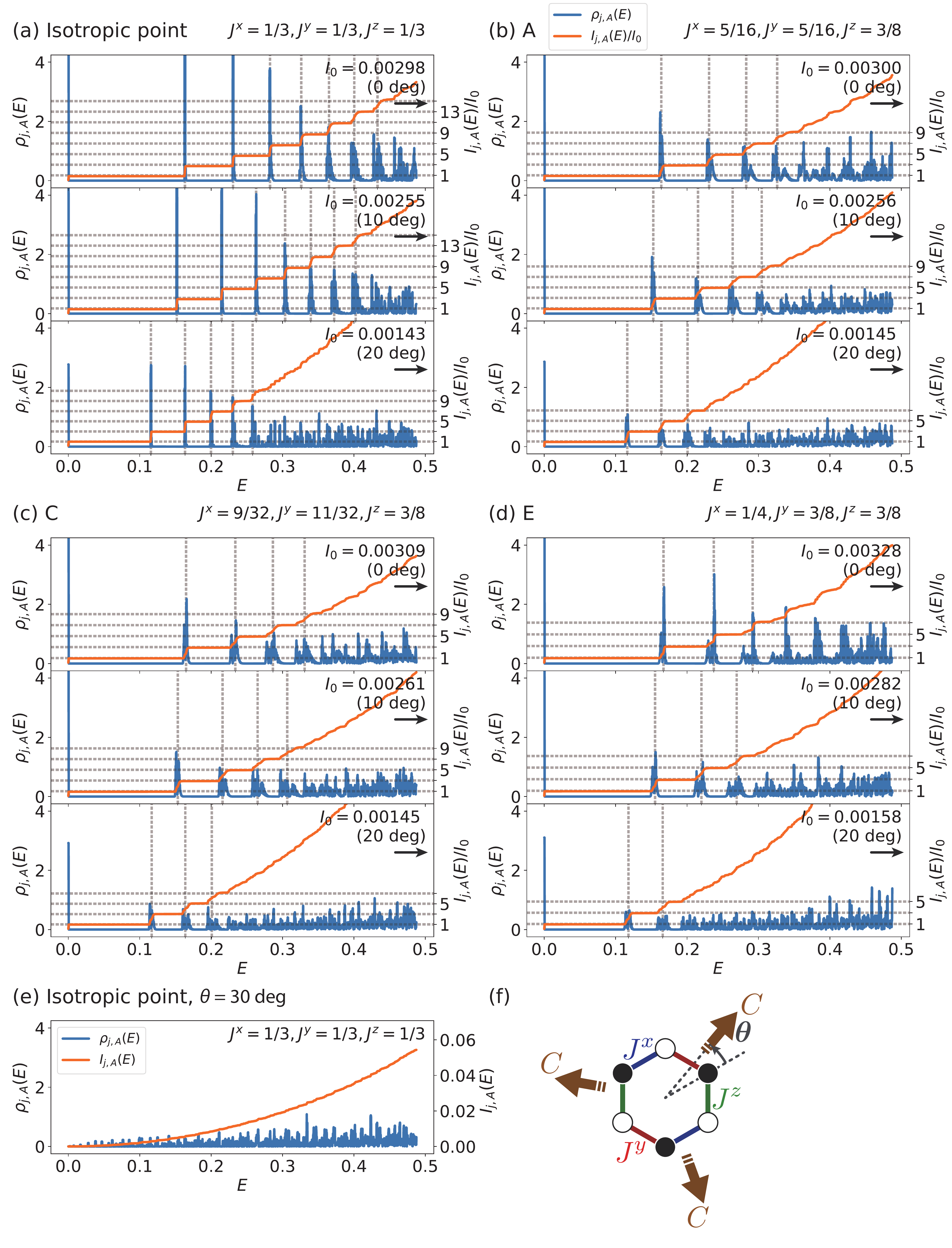}
\hspace{0pc}
\vspace{0pc}
\caption{(Color online)\label{rot} (a)-(e) Typical results for the LDOS, $\rho_{j, A}(E)$, and its integral value, $I_{j, A}(E)=\int_0^E \rho_{j, A}(E') dE'$, of the $A$ sublattice for various rotation angles, $\theta$. $I_{j, A}(E)$ is normalized by numerically evaluated $I_0$ to fit  $(2n+1) \; (n=0, 1, 2, \cdots)$. The coupling constants in (b)-(d) correspond to those in the middle panels of Figs. \ref{fig2}(b)-\ref{fig2}(d), respectively. Transverse dotted lines represent $(2n+1)$. We set $R=60$ and $C=1/300$. For $\theta=30^{\circ}$, $\rho_{j, A}(E)$ and $I_{j, A}(E)$ at the isotropic point are illustrated as representative in (e). (f) Deviation of triaxial strain expressed by $\theta$ from the bond direction. See also Fig.\ref{fig1}(b). }
\end{center}
\end{figure*}
We investigate the strain-induced Landau levels of Majorana fermions when the direction of triaxial strain deviates from the bond direction. The deviation is expressed by the rotation angle $\theta$ as shown in Fig. \ref{rot}(f). 
Figures \ref{rot}(a)-\ref{rot}(e) illustrate the typical results for the LDOS, $\rho_{j, A}(E)$, and its integral value, $I_{j, A}(E)=\int_0^E \rho_{j, A}(E') dE'$, of the A sublattice for $\theta=0-30^{\circ}$. Using the same way as described in Subsec. 2.2, we have confirmed that the ground state for $\theta \neq0$ is in the flux-free sector. 
For $\theta <30^{\circ}$, the integral value is well described by the relation, $I_{j, A}(E)/I_0=(2n+1)$, and $\rho_{j, A}(E)$ reaches a peak in the vicinity of the boundary between the neighboring plateaus of $I_{j, A}(E)$. Note that $I_0$ is evaluated so as to fit the numerical results.   
As $\theta$ increases, the peak energy $E_n \; (n\neq0)$ of $\rho_{j, A}(E)$ decreases, reflecting the decrease in the effective pseudomagnetic field.  
The weight of $\rho_{j, A}(E)$ is also reduced with increasing $\theta$. But, the pronounced multiple Landau levels are observed at least for $\theta \lessapprox 10^{\circ}$.  
In the vicinity of $\theta=30^{\circ}$, strain-induced Landau levels disappear [Fig. \ref{rot}(e)]. When $\theta$ is increased from $30^{\circ}$ to $60^{\circ}$, the energy $E_n \; (n\neq0)$ and the weight of  $\rho_{j, A}(E)$ increase with inverting the sublattice polarization: The peak at $E=0$ appears only at the $B$ sublattice in $30^{\circ}<\theta\leq60^{\circ}$. The strain-induced Landau levels at $\theta= 60^{\circ}$ are the same as those at $\theta= 0$ by inverting the sublattice polarization.      
The results indicate that the strain-induced Landau levels of Majorana fermions are robust against a slight rotation of triaxial strain from the bond direction.

\section{\label{sec:level4} Effective low-energy theory}
\subsection{\label{sec:level4-1}Landau levels of itinerant Majorana fermions by an analytical calculation}
We now discuss the low-energy states of the itinerant Majorana fermions on the triaxially-strained honeycomb lattice in the flux-free sector through an analytical calculation.
Following Refs. \cite{Poli,Ozawa2}, we adopt the effects of weak triaxial strain as $J^{\alpha}({\boldsymbol r})= J^{\alpha} \left[1+\tau\left({\boldsymbol r} \cdot {\boldsymbol r}^{\alpha}/ 3{a_0}^2 \right) \right]$ to eq. (\ref{Ham1}), where ${\boldsymbol r}^z=a_0(0, 1)$, $ {\boldsymbol r}^x=a_0(-\sqrt{3}/2, -1/2)$, and ${\boldsymbol r}^y=a_0(\sqrt{3}/2, -1/2)$ are vectors that connect the unstrained nearest-neighbor sites, as shown in Fig. \ref{fig1}(b), and $\tau$ controls the strain strength. 
Also in the effective low-energy theory, the coupling constants in the unstrained lattice satisfy $J^x+J^y+J^z=1$. 
In the anisotropically interacting system, the two inequivalent Dirac points move from the corners of the first Brillouin zone, $K$ and $K'$. We expand the off-diagonal elements of the sublattice-based $2\times2$ Hamiltonian around $K$ and $K'$, and take the zeroth and first order terms.

The pseudovector potential, $\boldsymbol{A}^{\xi}=\xi(A_x^{}, A_y^{})$, induced by triaxial strain is given as
\begin{equation}
%
\begin{split}
A_x= {|v_x|}^{-1}\Bigg[ \left(J^z-\frac{1}{2}J^x-\frac{1}{2}J^y\right)  + \left(J^x-J^y\right)\frac{x\tau}{4\sqrt{3}a_0}  \\
      +\left(J^z+\frac{1}{4}J^x+\frac{1}{4}J^y\right)\frac{y\tau}{3a_0} \Bigg], 
\end{split}
\end{equation}
\begin{equation}
A_y={|v_y|}^{-1}\frac{\sqrt{3}}{2} \left[ \left(J^x-J^y\right) -\left(J^x+J^y\right)\frac{x\tau}{2\sqrt{3}a_0} 
                -\left(J^x-J^y\right)\frac{y\tau}{6a_0} \right], 
\end{equation}
where $\xi=\pm 1$ for $K/K^{\prime}$ and
\begin{align*}
v_x^{\xi} &=\frac{\sqrt{3}a_0}{4\hbar}\left[\sqrt{3}\left(J^x+J^y\right)+i\xi\left(J^x-J^y\right)\right] \equiv |v_x|e^{i\xi\phi_x},  \\
v_y^{\xi} &=\frac{\sqrt{3}a_0}{4\hbar}\left[\frac{1}{\sqrt{3}}\left(3J^z+1\right)-i\xi\left(J^x-J^y\right)\right] \equiv |v_y|e^{i\xi\phi_y}.  
\end{align*}
The pseudomagnetic field, $\boldsymbol{B}^{\xi}=(0, 0, \xi  B_z)$, is given as $B_z^{}=\cos\phi_x\partial_xA_y-\cos\phi_y\partial_yA_x$. The Hamiltonian around $K$ and $K'$ reads 
%
\begin{eqnarray}
{\mathcal H}_{\xi}=\xi \sqrt{|v_xv_y|} 
\begin{pmatrix}
  0   & {\Pi^{\xi}_x}^{\ast}- i{\xi}{\Pi^{\xi}_y}^{\ast} \\
  \Pi^{\xi}_x+ i{\xi}\Pi^{\xi}_y & 0 \\ 
\end{pmatrix},
\label{Ham2}
\end{eqnarray}
where $\Pi^{\xi}_x=\sqrt{|v_x/v_y|} \left[e^{i\xi \phi_x}p_x+\xi A_x^{} \right]$ and $\Pi^{\xi}_y=\sqrt{|v_y/v_x|} \left[e^{i\xi \phi_y}p_y+\xi A_y^{} \right]$. 
By defining the annihilation operator as $a_{\xi} =(l_B/\sqrt{2}\hbar) \left(\Pi^{\xi}_x+i\xi\Pi^{\xi}_y \right)$
with the magnetic length ${l_B}=\sqrt{\hbar/|B_z|}$, the eigenenergy is obtained as $E_n=\left(2\sqrt{2}\hbar/l_B \right) \sqrt{|v_xv_y|} \sqrt{n}, \; (n=0, 1, 2, \cdots)$. 
The $n=0$ eigenstate is $|\Psi^{}_0\rangle=\left(|\psi_0\rangle, 0 \right)^T$ for both $K$ and $K'$, indicating that the eigenstate at $n=0$ is nonzero only on the $A$ sublattice (sublattice polarization), while the $n\neq0$ eigenstates are $|\Psi^{\xi}_n\rangle=\left(|\psi_{|\xi n|}\rangle, \xi|\psi_{|\xi n|-1}\rangle \right)^T$ for $K/K'$. Here, the spinor components satisfy $a_{\xi}^{\dagger}a_{\xi}|\psi_{|\xi n|}\rangle=n|\psi_{|\xi n|}\rangle$. These eigenenergy and eigenstate features agree with the numerical results, providing the evidence of the emergent Landau levels of Majorana fermions.   

We next discuss the effects of the direction of triaxial strain. Rotating ${\boldsymbol r}^{\alpha}$ by the angle $\theta$, we obtain the $\theta$ dependent coupling constant $J^{\alpha}({\boldsymbol r})$, which yields the $\theta$ dependent pseudovector potential. The pseudomagnetic field thus obtained, $B_z$, is proportional to $\cos(3\theta)$. This result indicates that the magnitude of the pseudomagnetic field decreases according to $\cos(3\theta)$ with increasing the rotation angle $\theta$. At $\theta=30^{\circ}$, the pseudomagnetic field disappears. This $\theta$ dependence explains the numerical results illustrated in Figs. \ref{rot}(a)-\ref{rot}(e) and the inverted sublattice polarization in $30^{\circ}<\theta\leq60^{\circ}$.

\subsection{\label{sec:level4-2}Relation between the strain strengths in the numerical and analytical calculations}
\begin{figure}[thb]
\begin{center}
\includegraphics[width=0.9\hsize]{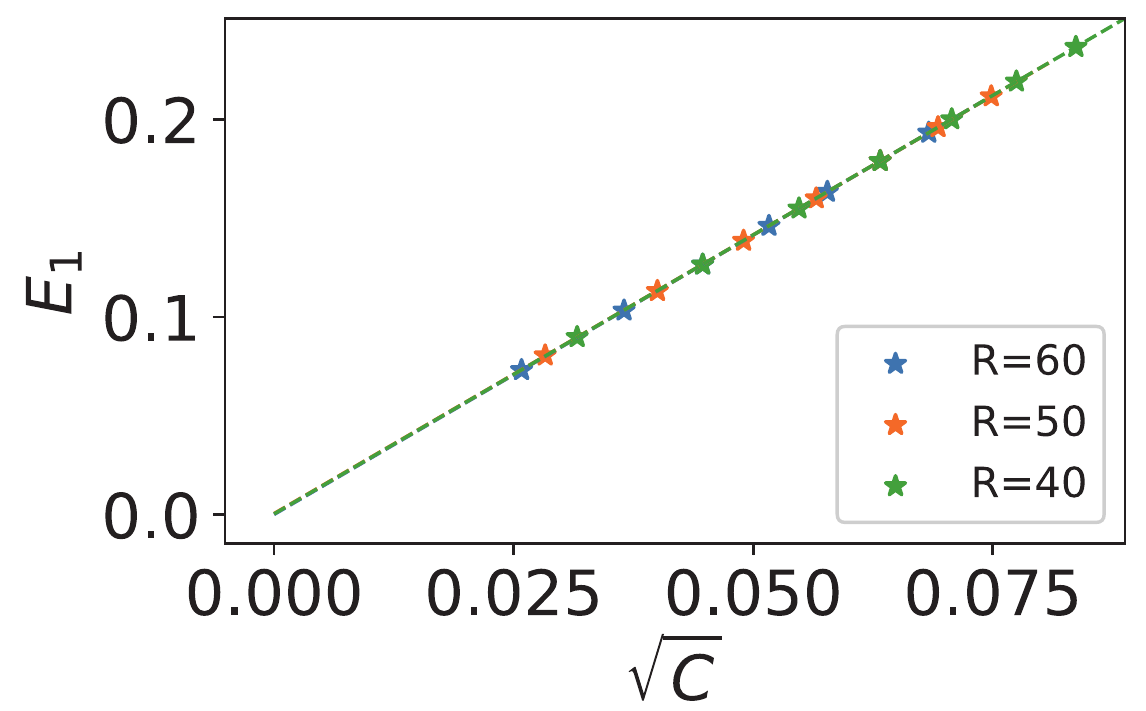}
\hspace{0pc}
\vspace{0pc}
\caption{(Color online)\label{fig5} Peak energy, $E_1$, of  $\rho_{j, A}(E)$ as a function of  $\sqrt{C}$ at the isotropically interacting system obtained in the numerical calculation. $R=$40, 50, and $60$ are set. Corresponding $C$ is obtained for the fixed $CR=$1/25,  2/25, 3/25, 4/25, 5/25, 6/25, and 7/25, respectively.}
\end{center}
\end{figure}
\begin{table*}[htb]
\begin{center}
 \caption{\label{table1} Parameters $a$ and $b$ to fit the coefficient ($E_1$) of $E_n \propto \sqrt{n}$ using a linear regression, $E_1=a\sqrt{8C}+b$. The coefficients ($E_1$) are obtained by the numerical calculation for $R=$40, 50, and 60 at the fixed $CR=$1/25,  2/25, 3/25, 4/25, 5/25, 6/25, and 7/25, respectively.
The coupling constants are for the isotropically interacting system and for the black dots closest to the isotropically interacting point on the lines A-E in Fig. \ref{PD}. 
}
  \begin{tabular}{ccccccc}
  \hline  
  $(J^x, J^y, J^z)$&$(\frac{1}{3}, \frac{1}{3}, \frac{1}{3})$&A $(\frac{31}{96}, \frac{31}{96}, \frac{17}{48})$ & B $(\frac{121}{384}, \frac{127}{384}, \frac{17}{48})$ & C $(\frac{59}{192}, \frac{65}{192}, \frac{17}{48})$ & D $(\frac{115}{384}, \frac{133}{384}, \frac{17}{48})$ & E $(\frac{7}{24}, \frac{17}{48}, \frac{17}{48})$\rule[-2mm]{0mm}{5mm}\\ \hline
 $a$&$0.9995$&$0.9997$&$1.0011$&$1.0042$&$1.0072$&$1.0099$\\
$\frac{a}{0.9995}$&1&1.0002&1.0002&1.0005&1.0077&1.0104\\
 $b$&$9.0367\times10^{-5}$&$1.4873\times10^{-4}$&$2.5185\times10^{-5}$&$-2.1382\times10^{-4}$&$-3.8310\times10^{-4}$&$-4.6970\times10^{-4}$\\
  \hline 
  \end{tabular}
\end{center}
\end{table*}
We investigate the relation between the control parameters of the strain strength, $C$ and $\tau$, in the numerical and analytical calculations, respectively. To this end, we evaluate the coefficient of $E_n\propto \sqrt{n}$ obtained in the numerical calculation for $R=$40, 50, and 60 at the seven values of the fixed $CR$ within $CR<0.3$. 
Figure \ref{fig5} illustrates the coefficient ($E_1$) as a function of $\sqrt{C}$ in the isotropically interacting system. The data follows the line $E_1=a\sqrt{8C}+b$ with $a\approx0.9995$ and $b\approx9.0367\times10^{-5}$, indicating that the relation, $E_n =\sqrt{8Cn}$, is deduced within our numerical accuracy. 
The coefficient $E_1$ in the system closest to the isotropically interacting point on each line A-E is evaluated in the same way. The evaluated $a$ and $b$ are summarized in Table I, indicating that $a\approx1.0$ and $b\approx 0$, and thus $E_n \approx\sqrt{8Cn}$.

The coefficient of $E_n \propto \sqrt{n}$ in the effective low-energy theory, $\left(2\sqrt{2}\hbar/l_B \right) \sqrt{|v_xv_y|}$, is expressed by using $\tau$ and one of the three coupling constants, leading to the expression for $E_n$ as 
\begin{eqnarray}
E_n=\left[\left(J^z-1\right)^2+\left(J^z+\frac{1}{3}\right)^2 \right]^{\frac{1}{2}}\sqrt{\frac{3\tau n}{2}}.
\label{Coeff}
\end{eqnarray}
The eigenenergy in the isotropically interacting system is obtained as $E_n=2\sqrt{\tau n/3}$. Comparing $E_n$ in the numerical and analytical calculations for the isotropically interacting system, the relation, $\tau=6C$, is derived. This relation is consistent with that derived by comparing the pseudomagnetic field in the numerical \cite{Settnes} and analytical calculations for the isotropically interacting system, where $|B_z|=4C\hbar/{a_0}^2$ and $2\hbar\tau/(3{a_0}^2)$, respectively.
Since $J^z$'s in the anisotropically interacting system in Table I are the same, their coefficients of $\sqrt{\tau}$ take the same value, $\sqrt{1025/768}$, according to Eq. (\ref{Coeff}). The ratio of this coefficient to that in the isotropically interacting system is $1.0005$, which is close to that denoted in Table I. Therefore, the relation $\tau=6C$ is approximately satisfied in the anisotropically interacting systems denoted in Table I, allowing a quantitative comparison between the numerical and analytical calculations.

\section{\label{sec:level5} Summary}
We have investigated the energy structure of an anisotropically interacting Kitaev model under triaxial strain. 
The emergence of the strain-induced Landau levels of itinerant Majorana fermions in the anisotropically interacting Kitaev model has been confirmed by the numerical and analytical calculations. These Landau levels are stable, when the direction of triaxial strain deviates slightly from the bond direction. 
It was shown that Raman spectroscopy can detect the features of the strain-induced Landau levels of Majorana fermions \cite{Perreault}. Scanning tunneling microscopy (STM) may also be promising \cite{Udagawa,Bauer}. Developing a theory of STM based on experimental systems is a future study.
The fabrication of $\alpha$-${\rm RuCl_3}$ thin film has been studied actively in recent years \cite{Weber,Gronke,Zhou,Yang}. 
If the method for generating triaxial strain in graphene nanobubbles \cite{Levy,Lu} can be applied to the $\alpha$-${\rm RuCl_3}$ thin film, it may be possible to generate triaxial strain in the $\alpha$-${\rm RuCl_3}$ thin film as well. This situation favors experiments for investigating the strain-induced Landau levels of Majorana fermions in Kitaev candidate material $\alpha$-${\rm RuCl_3}$. 
We hope that our results contribute to such studies.

\begin{acknowledgments}
We would like to thank T. Suzuki and R. Taniguchi for valuable discussions.  
This work was supported by JSPS KAKENHI Grant Number JP19K03721. 
\end{acknowledgments}

\end{document}